\documentclass[12pt]{article}

\newcommand{\x}{arXiv:}
\newcommand{\m}{\mathrm}
\newcommand{\be}{\begin{equation}}
\newcommand{\ee}{\end{equation}}
\newcommand{\ba}{\begin{eqnarray}}
\newcommand{\ea}{\end{eqnarray}}
\newcommand{\dif}{\mathrm{d}}

\usepackage{graphicx}
\usepackage{amssymb}
\usepackage{amsmath}
\usepackage[T1]{fontenc} %for \boldsymbol
\usepackage[ansinew]{inputenc} %for \boldsymbol
\usepackage[nosort]{cite}
\newcommand{\inbar}{\vrule height1.57ex width.4pt depth0pt}
\newcommand{\SW}{\relax{\hbox{$\ \inbar\kern-.285em{\rm S}$}}}

\begin{document}
\thispagestyle{empty}
\begin{center}

\null \vskip-1truecm \vskip2truecm

{\Large{\bf \textsf{Field Theories Without a Holographic Dual }}}

{\Large{\bf \textsf{}}}

{\large{\bf \textsf{}}}

{\large{\bf \textsf{}}}

\vskip1truecm

{\large \textsf{Brett McInnes
}}

\vskip0.1truecm

\textsf{\\ National
  University of Singapore}
  \vskip1.2truecm
\textsf{email: matmcinn@nus.edu.sg}\\

\end{center}
\vskip1truecm \centerline{\textsf{ABSTRACT}} \baselineskip=15pt
\medskip
In applying the gauge-gravity duality to the quark-gluon plasma, one models the plasma using a particular kind of field theory with specified values of the temperature, magnetic field, and so forth. One then assumes that the bulk, an asymptotically AdS black hole spacetime with properties chosen to match those of the boundary field theory, can be embedded in string theory. But this is not always the case: there are field theories with no bulk dual. The question is whether these theories might include those used to study the actual plasmas produced at such facilities as the RHIC experiment or the relevant experiments at the LHC. We argue that, \emph{provided} that due care is taken to include the effects of the angular momentum associated with the magnetic fields experienced by the plasmas produced by peripheral collisions, the existence of the dual can be established for the RHIC plasmas. In the case of the LHC plasmas, the situation is much more doubtful.

\newpage
\addtocounter{section}{1}
\section* {\large{\textsf{1. String Theory in the Bulk}}}
Attempts to apply gauge-gravity duality to the Quark-Gluon Plasma \cite{kn:youngman,kn:gubser,kn:janik} have always to reckon with the fact that QCD itself certainly does not have any \emph{known} description of this kind; if it has one at all, the dual is, to put it very mildly, not simple \cite{kn:karch}. Instead, one confines attention to greatly simplified versions of string theory in the bulk ---$\,$ the string coupling should be very small, the string length scale should be small relative to the bulk curvature length scale ---$\,$ and accepts that the corresponding simplified boundary field theories differ, in important ways, from QCD. The hope is nevertheless that all of these theories, including QCD, have some universal features in common \cite{kn:namit}.

In practice, one studies a field theory of this general type, with prescribed parameters (temperature, chemical potential, and so on) chosen to match those of the QGP, and then constructs the appropriate bulk geometry. When using this simplification, one should however bear in mind that the true bulk physics is still \emph{string} physics \cite{kn:mateos}. The procedure implicitly assumes that arbitrary boundary field theories with specific prescribed parameter values are dual to stringy bulk configurations that actually exist. \emph{But this is a very non-trivial assumption}.
For it is clear that, for a holographic description be be possible at all, the bulk physics must be extremely strongly constrained; there must be severe additional restrictions, beyond those imposed by classical General Relativity, if it is to be fully equivalent to a lower-dimensional dual. In particular, many candidate bulk geometries, supposedly dual to a constructed boundary field theory, must in fact be mathematically inconsistent when embedded in string theory.

In summary: not every boundary field theory system can have a gravitational dual; one will find in some cases that the field theory is ``dual'' to a system which does not actually exist in full string theory, even if it \emph{appears} to do so in an incautious application of the standard holographic procedure.

All this is of considerable theoretical interest, since, as we shall see, recent advances make it rather easy to exhibit explicit examples of this phenomenon. More importantly, however, it prompts the question: are there field theory systems, corresponding (as above) to \emph{physical, experimentally attainable QGP states}, such that the purported bulk dual spacetime simply does not exist in a full string-theoretic treatment? Are there, in short, actual plasmas with no holographic description?

Our objective in this work is to bring together some recent important developments in string theory \cite{kn:ferrari1,kn:ferrari2,kn:ferrari3,kn:ferrari4} with new phenomenological findings (particularly \cite{kn:holliday}), in order to argue that there is a real possibility that such systems might indeed exist; the QGP arising in certain heavy-ion collisions (involving extremely intense magnetic fields) corresponds to a field-theory configuration which appears to be dual to a bulk system that is not mathematically consistent in string theory. That is, the holographic duals of certain specific quark plasmas indeed (apparently) do not exist.

Remarkably, the boundary between plasmas with a consistent holographic description and those (possibly) without one lies between the regimes explored by the main experimental facilities: on the one hand, the plasmas which are the concern of the RHIC experiment and the allied beam energy scans do have such a description, while, on the other, the QGP produced in certain peripheral heavy-ion collisions at the LHC (and potentially in future facilities such as the Future Circular Collider) apparently do not. We will see, in fact, that it is not trivial to establish the existence of such a description even in the case of certain RHIC plasmas; this can be done, but only by explicitly including certain effects (the shearing and vorticity of the plasma) associated with the magnetic fields.

\addtocounter{section}{1}
\section* {\large{\textsf{2. The Consistency Condition vs. Magnetic Fields}}}

The argument proceeds as follows. In \cite{kn:ferrari1,kn:ferrari2,kn:ferrari3,kn:ferrari4} the authors argue that, in an extremely broad class of dual bulk-boundary pairs, a mathematically consistent string-theoretic bulk must satisfy a simple relation between the (on-shell) Euclidean spacetime action and the (on-shell) action of probes such as branes. This is argued to be related to very deep and general thermodynamic aspects of gravitation, and has been confirmed by an impressive array of highly non-trivial checks \cite{kn:ferrari4}.

This relation takes very explicit forms in certain special cases. One example ---$\,$ we stress that it is but \emph{one} of the many constraints implied by the consistency condition, though it will be the main focus of the present work ---$\,$ can be stated as follows. The requirement is that for each $d\,-\,$dimensional hypersurface $\Sigma$ embedded in, and homologous to the conformal boundary of, a $(d+1)\,-\,$dimensional (Euclidean) bulk, the area $A(\Sigma)$ and the volume $V(M_{\Sigma})$ enclosed by $\Sigma$ must satisfy the ``isoperimetric inequality''
\begin{equation}\label{A}
\mathfrak{S^{\m{E}}}\;\equiv\;A(\Sigma)\;- \;{d \over L}V(M_{\Sigma}) \;\geqslant \;0,
\end{equation}
where, henceforth, $L$ denotes the asymptotic AdS curvature scale, and the superscript ``E'' denotes a Euclidean quantity. This condition is a very subtle global restriction on the bulk geometry: it demands that the areas of these distinguished surfaces should dominate their (suitably normalized) volume \emph{throughout} the bulk. It is satisfied in Euclidean $AdS_{d+1}$ (in fact the left side vanishes identically if one foliates the relevant submanifold of Euclidean $AdS_{d+1}$ by planes perpendicular to the radial direction), and it is satisfied in many other Euclidean asymptotically AdS bulk spacetimes. \emph{But it is not satisfied in all}. In the latter case, the conclusion is that the proposed bulk configuration does not exist within string theory. The physical interpretation is that \emph{some unexpected condition must be imposed on the boundary field theory} if it is to have a genuine bulk dual.

We stress that a failure to satisfy (\ref{A}) in a concrete physical application of holography would be a serious matter indeed. This condition has a Lorentzian counterpart which, beginning with \cite{kn:seiberg,kn:wittenyau}, has been studied extensively. When that condition fails, the result is an instability. But for the systems in which we are interested here, which have extremely short lifetimes, it is not clear that such an instability is relevant, since there might not be sufficient time for it to evolve before the plasma ceases to exist in any case. In the Euclidean case, such questions do not arise: a violation of (\ref{A}) simply means that the formalism breaks down and holography cannot be used.

Some important cases where this inequality \emph{is} satisfied were studied in \cite{kn:ferrari3}. There, the results of \cite{kn:lee,kn:wang,kn:wang2} (see also \cite{kn:galloway}) were used to demonstrate that (\ref{A}) holds throughout the bulk when the Yamabe invariant of the boundary manifold is non-negative and the bulk is an \emph{Einstein} manifold. Thus for example there is no difficulty in embedding Euclidean AdS-Kerr geometry in string theory (the boundary in that case being a product of a circle with a sphere, hence having non-negative Yamabe invariant).

However ---$\,$ and this is a key issue ---$\,$ one is often interested in a bulk geometry that is \emph{not} an Einstein manifold. For example, to treat a quark-gluon plasma at non-zero baryonic chemical potential, one needs to consider an electric field in the bulk, and this deforms the latter away from being Einstein. Similarly, there has recently been intense interest in the extreme magnetic fields associated with the QGP produced in peripheral collisions
\cite{kn:skokov,kn:tuchin,kn:magnet,kn:review} at heavy-ion facilities. The holographic treatment of this system, which will be our main focus in this work, requires a magnetic field in the bulk \cite{kn:hartkov,kn:dyon,kn:82,kn:83,kn:84,kn:86,kn:87}. The bulk metric again ceases to be Einstein when the back-reaction from the magnetic field is taken into account.

Thus, it is not clear that the inequality (\ref{A}) must always hold in these cases. It will hold for small deviations away from the Einstein condition, but, as was shown in \cite{kn:83}, not always under more extreme conditions; not, in particular, when the magnetic field is very strong.

The point, however, is that experiments involving heavy ion collisions can give rise to plasmas immersed in magnetic fields which \emph{are} ``very strong'', so we need to consider the form taken by (\ref{A}) in this case. It was shown in \cite{kn:82} (in the case of approximately zero baryonic chemical potential) that (\ref{A}), evaluated for a suitable version of the magnetic AdS  Reissner-Nordstr\"om geometry, can be translated, through holography, to a surprisingly simple relation between magnetic field $B$ experienced by the boundary field theory and its temperature\footnote{As is well known, in string theory one expects $T$ itself to be bounded above by the Hagedorn temperature (see \cite{kn:hagedorn} for possible consequences for holography), so in a sense this inequality is the analogous one for magnetic fields.} $T$: the dual version of (\ref{A}) is just
\begin{equation}\label{AA}
B\;\leq \;2\pi^{3/2}T^2\;\approx \; 11.14 \times T^2.
\end{equation}

We see that, to the extent that the actual QGP can be adequately described by its temperature and the magnetic field it experiences, the internal mathematical consistency of the bulk theory has explicit consequences for observable parameters. Still more remarkable, as we shall now show, is that the actual values come close to saturating this inequality ---$\,$ and in some cases may actually violate it.

\addtocounter{section}{1}
\section* {\large{\textsf{3. The QGP at the RHIC and in the Beam Energy Scans}}}
The classic study of Skokov et al. \cite{kn:skokov} considered the maximal\footnote{Of course, any given heavy-ion beam produces plasmas with magnetic fields of varying intensity, and various temperatures, depending on the impact parameter and other variables. Henceforth, to avoid tedious repetition, we always mean maximal fields whenever magnetic fields are mentioned.} magnetic fields arising in peripheral collisions at the RHIC facility: the estimate is $eB \approx m_{\pi}^2$, where $e$ is the electron charge and where $m_{\pi}$ is the pion mass. In natural units, with $m_{\pi} \approx 0.71$ fm$^{-1}$, this means $B \approx 1.67$ fm$^{-2}$. Typical temperatures in these collisions (for which $\mu_B = 0$ is indeed a good approximation) are around 220 MeV, or $T \approx 1.12$ fm$^{-1}$. The right side of (\ref{AA}) is then $\approx 13.97$ fm$^{-2}$. Thus (\ref{AA}) is satisfied, and consequently so is (\ref{A}); even if one trusts holography only up to factors of around 2 \cite{kn:karch}\cite{kn:mateos}, it seems that all is well. That was our conclusion in \cite{kn:84}.

However, the data for the more recent LHC experiments are less comforting. In \cite{kn:skokov}, the estimate for collisions at the LHC \cite{kn:armesto} was that $B$ should be around 15 times larger than at the RHIC, whereas the temperature only increases to about 300 MeV: the left side of (\ref{AA}) is then $\approx 25.1$ fm$^{-2}$, while the right side is $\approx 25.8$ fm$^{-2}$. Although it is true that quantities like ``the temperature'' of the plasma in these very extreme conditions have to be interpreted carefully, one now feels somewhat less confident that all is indeed well: \emph{as the impact energy goes up, the magnetic field rises much more quickly than the temperature}. However, let us delay consideration of that case, which involves additional subtleties; until further notice we will focus on the RHIC and the associated beam energy scan experiments.

Even in the case of the RHIC, there is a problem: since the publication of \cite{kn:skokov}, estimates for $B$ have been moving rather sharply upward, without a corresponding increase in values for $T$. Recent discussions (for example \cite{kn:fivetimes}) have led to estimates of the maximal RHIC magnetic field of around $eB \approx 5 \times m_{\pi}^2$, putting $B \approx 8.35$ fm$^{-2}$, still less than, but uncomfortably close to, the right side of the inequality (as above, $\approx 13.97$ fm$^{-2}$). Another relevant theme here is the distinction between \emph{average} fields computed over many collisions, and ``event-by-event'' analyses \cite{kn:bzdak}, which again lead to larger estimates \cite{kn:shipu}, as high as $eB \approx 10 \times m_{\pi}^2$. The event-by-event value is the relevant one here; but such a field, with $B \approx 16.64$ fm$^{-2}$, clearly violates\footnote{Still more recently, an interesting investigation \cite{kn:veryhigh} has (``optimistically'') considered still larger values for the maximal $B$, around $eB \approx (500 \m{MeV})^2 \approx 12.8 \times m_{\pi}^2$ or $B \approx 21$ fm$^{-2}$. Some lattice investigations have contemplated even more extreme situations, as for example \cite{kn:gergely}; the magnetic fields considered there ($eB = 3.25$ GeV$^2$) violate (\ref{AA}) by well over an order of magnitude, even for LHC temperatures. Similar comments apply to other investigations of various forms of ``catalysis'' with ultra-high fields \cite{kn:cat}. While it is not claimed that these theoretical considerations necessarily correspond to any actually realisable physical system, it is presumably important to understand that the existence of a holographic dual is open to doubt in these cases.} the inequality (\ref{AA}). There are similar upward revisions for the LHC case, as we will discuss later.

In an important recent development, Holliday and Tuchin \cite{kn:holliday} have provided a strong theoretical basis for these higher estimates. They observe that previous calculations of magnetic fields in these circumstances have neglected the quantum diffusion of the nucleon wave function, and that, when this is taken into account, the computed values of the fields are considerably larger than earlier estimates.

In short, values for $B$ and $T$ violating (\ref{AA}), say $eB \approx 10 \times m_{\pi}^2$ and $T \approx 220$ MeV in the case of the RHIC plasma, must now be taken very seriously. It therefore \emph{seems} that some of the plasmas studied in peripheral collisions at the RHIC experiment correspond to field theories which do not have a holographic dual: the bulk spacetime is apparently acceptable classically, but not in string theory.

Here we wish to point out that there is an extremely natural and simple way to avoid this conclusion: our discussion above neglects the effects of a fundamentally important physical aspect of these plasmas, to wit, \emph{they have an enormous density of angular momentum.}

It has in fact long been known that, precisely in the case of peripheral collisions, a very large amount of angular momentum is transferred to the plasma \cite{kn:liang,kn:bec,kn:huang,kn:KelvinHelm,kn:viscous,kn:csernairecent1,kn:csernairecent2,kn:nagy,kn:nacs,kn:yin,kn:deng,kn:vortical}; this arises from exactly the same circumstances that give rise to the magnetic field, and the two effects are inseparable. From the field theory point of view, it is not obvious why this is relevant. But from a holographic point of view, its relevance is immediately clear: angular momentum in the boundary theory must be associated with a bulk black hole endowed with angular momentum, and this of course has a strong effect on the bulk spacetime geometry ---$\,$ we need to use a suitable generalization of the Kerr-Newman, instead of Reissner-Nordstr\"om, geometry. The presence of angular momentum, by changing the bulk geometry, affects $\mathfrak{S^{\m{E}}}$, and consequently it means that we must re-consider whether (\ref{A}) is indeed violated by field theories modelling actual quark plasmas produced in peripheral collisions.

As is notorious, black holes with angular momentum are very complex objects, and it is by no means clear that taking this effect into account will save the situation; it might just as easily make it worse. We need a detailed investigation. As preparation for that, we briefly review the derivation of (\ref{AA}), with a view to its subsequent generalization.

\addtocounter{section}{1}
\section* {\large{\textsf{4. A Review of the Magnetic Bound Without Angular Momentum}}}
The relevant bulk black hole \cite{kn:dyon} in the absence of angular momentum is a Euclidean dyonic asymptotically AdS four-dimensional Reissner-Nordstr\"om black hole with a flat event horizon; the metric is
\begin{eqnarray}\label{B}
g^E(\m{AdSdyRN^{0}_{4})} & = &  \Bigg[{r^2\over L^2}\;-\;{8\pi M^*\over r}+{4\pi (-\,Q^{*2}+P^{*2})\over r^2}\Bigg]\m{d}t^2\; \nonumber \\
& &  + \;{\m{d}r^2\over {\dfrac{r^2}{L^2}}\;-\;{\dfrac{8\pi M^*}{r}}+{\dfrac{4\pi (-\,Q^{*2}+P^{*2})}{r^2}}} \;+\;r^2\Big[\m{d}\psi^2\;+\;\m{d}\zeta^2\Big];
\end{eqnarray}
here $M^*$, $Q^*$, and $P^*$ are parameters related to the mass, electric charge, and magnetic charge per unit horizon area (see \cite{kn:77}), $t$ and $r$ have the expected interpretations, and $\psi$ and $\zeta$ are dimensionless coordinates on the plane or torus transverse to the radial direction. (At infinity, in the Lorentzian version, they define, respectively, the standard coordinates $x$ and $z$ in the reaction plane of a heavy-ion collision; here $z$ is the axis of the collision.)

Keep in mind that this metric is \emph{not} an Einstein metric: this simple fact is the core of the problem.

In the usual way, the Lorentzian versions of the black hole parameters have ``holographic'' physical interpretations as quantities describing the dual field theory: its temperature $T$, baryonic chemical potential $\mu_B$, and its associated magnetic field $B$. The relations (see \cite{kn:83}) are
\begin{equation}\label{C}
T\;=\;{r_h\over \pi L^2}\;-\;{2M^*\over r_h^2}.
\end{equation}
\begin{equation}\label{D}
\mu_B\;=\;{3Q^*\over r_hL},
\end{equation}
\begin{equation}\label{E}
B \;=\; P^*/L^3,
\end{equation}
where $r_h$ denotes the value of the radial coordinate at the \emph{Lorentzian} event horizon; it is related to the black hole parameters in the usual way:
\begin{equation}\label{F}
{r_h^2\over L^2}\;-\;{8\pi M^*\over r_h}+{4\pi (Q^{*2}+P^{*2})\over r_h^2}\;=\;0.
\end{equation}
These four equations provide the ``holographic dictionary'' in this case: for example, given the field theory parameters $T$, $\mu_B$, $B$, together with $L$ (which, for reasons to be explained, we take to be around 10 femtometres), one can solve for the four black hole parameters $M^*$, $Q^*$, $P^*$, and $r_h$.

For this black hole, $\mathfrak{S^{\m{E}}}$ is readily computed explicitly: it takes the form (up to an overall positive constant factor) of a function of $r$, given by
\begin{equation}\label{G}
\mathfrak{S^{\m{E}}}(\m{AdSdyRN^{0}_{4}})(r)\;=\; { \left (-8\pi M^* + {\dfrac{4\pi (-\,Q^{*2} + P^{*2})}{r}} \right )/L\over 1+\sqrt{1-{\dfrac{8\pi M^*L^2}{r^3}}+{\dfrac{4\pi (-\,Q^{*2} + P^{*2})L^2}{r^4}}}}+{(r^{\m{E}}_h)^3\over L^3}.
\end{equation}
Here $r^{\m{E}}_h$ locates the ``Euclidean event horizon'', which is essentially just the origin of coordinates in the Euclidean $r-t$ plane; it is given by solving an equation identical to equation (\ref{F}) except that the sign of $Q^{*2}$ is reversed in passing to the Euclidean case. Of course, $\mathfrak{S^{\m{E}}}(\m{AdSdyRN^{0}_{4})}(r)$ vanishes at the Euclidean event horizon, since the area and volume are both zero there, and it is not difficult to see that it is positive nearby. But it need not be positive farther away from the origin.

In fact, simple modifications of the calculations in \cite{kn:83} (which discussed the Lorentzian case) show that this function is never negative if and only if
\begin{equation}\label{H}
4\pi (P^{*2}\,-\,Q^{*2})L^2 \;\leqslant \;(r^{\m{E}}_h)^4.
\end{equation}

Because $r^{\m{E}}_h$ is a function of all of the other variables, this relation is more complex than it looks; however, the following statements are useful heuristic guidelines\footnote{We do not have formal proofs of the following statements, but in each case one can construct convincing plausibility arguments. For example, in the first case, note that these black holes, unlike ``small'' AdS black holes with spherical event horizons, always have a positive specific heat, so one can expect the entropy, and therefore the Lorentzian horizon radius $r_h$, always to increase with the temperature. (This does not follow from equation (\ref{C}), since one has no justification for fixing $M^*$ in this case.) Since it easy to see that, for fixed values of the black hole parameters, $r_h \leq r^{\m{E}}_h$, one can expect that the right side will also increase with $T$. All of these statements are supported by numerical evidence.}.

$\bullet$ The effect of increasing the temperature will be to increase the right side of (\ref{H}). That is, high temperatures have a favourable effect from the point of view of maintaining the consistency condition: this is reflected in (\ref{AA}) in the special case of zero baryonic chemical potential.

$\bullet$ The effect of increasing the baryonic chemical potential, which is related to the electric charge, is likewise favourable.

$\bullet$ The effect of increasing the magnetic field is unfavourable; again, see (\ref{AA}) for the special case $\mu_B = 0$.

Thus the danger of having a string-theoretically inconsistent bulk is at its greatest when $B$ is large while $\mu_B$ is small (or, in theory, if $T$ is small).

Now, in fact, peripheral heavy-ion collisions at the RHIC and (even more so) at the LHC do involve small $\mu_B$ and large $B$; so these experiments do potentially explore precisely the parameter domain in which the consistency condition is most at risk. To be precise, the dangerous region is characterised by putting (from equation (\ref{D})) $Q^* = 0$ in (\ref{H}); notice that this implies that $r_h = r^{\m{E}}_h$ in this case. A straightforward calculation \cite{kn:82} using equations (\ref{C}), (\ref{E}), and (\ref{F}) to convert from black hole parameters to boundary parameters then shows that (\ref{H}) is equivalent to the inequality (\ref{AA}) when the baryonic chemical potential can be neglected. This, then, is the very concrete form taken by the consistency condition if we attempt to use the gauge-gravity duality to describe these particular plasmas.

To see all this explicitly, let us consider the case of a plasma at typical RHIC temperatures (and approximately zero $\mu_B$), $T \approx 1.12$ fm$^{-1}$, and let us take a high, but (as we argued above) by no means outlandish, estimate for the magnetic field: $eB \approx 10 \times m_{\pi}^2$, or $B \approx 16.64$ fm$^{-2}$. These parameter values violate (\ref{AA}), and indeed the graph of $\mathfrak{S^{\m{E}}}(\m{AdSdyRN^{0}_{4}})(r)$ in this case (Figure 1) shows that it does in fact become negative sufficiently far from the Euclidean horizon. (Here, and in all of our graphs henceforth, the vertical axis has been scaled for convenience, so the values on that axis have no significance.)
\begin{figure}[!h]
\centering
\includegraphics[width=0.65\textwidth]{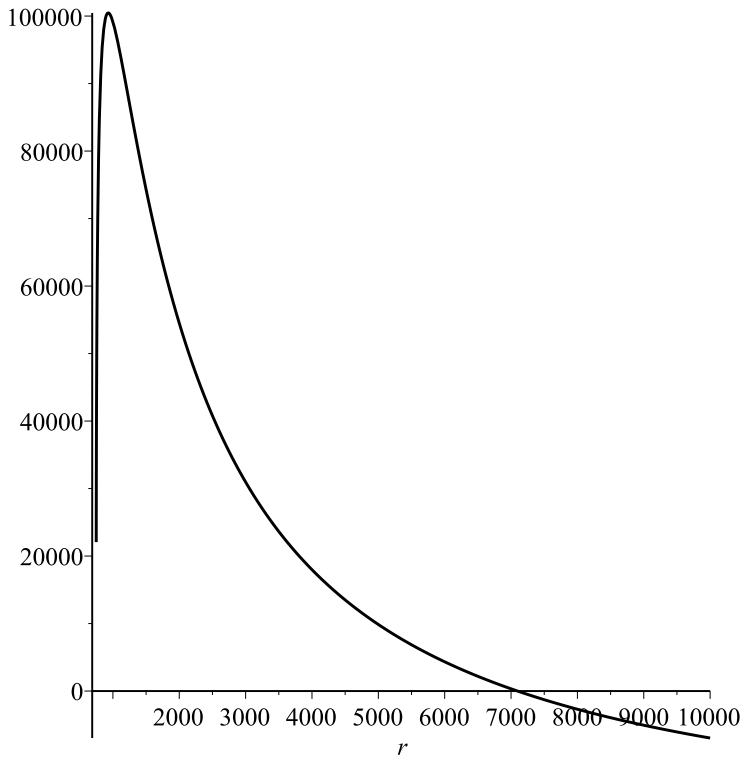}
\caption{$\mathfrak{S^{\m{E}}}(\m{AdSdyRN^{0}_{4}})(r)$, $T \approx 1.12$ fm$^{-1}$, $\mu_B = 0$, $B \approx 16.64$ fm$^{-2}$.}
\end{figure}
Thus we certainly have a problem in the RHIC case, with these parameter values.

On the other hand, the other potentially dangerous parameter domain is that of small $T$. Experimentally, this is the domain of the beam energy scan experiments currently under way or planned \cite{kn:STAR,kn:shine,kn:nica,kn:fair,kn:BEAM,kn:luo}. However, the temperature of the plasma cannot of course be arbitrarily low, since the plasma hadronizes as temperatures are lowered, either through a crossover, or through a phase transition at lower temperatures than the crossover. These lower temperatures occur, however, in conjunction with lower maximal values of the magnetic field and with higher values of the baryonic chemical potential, both of which are favourable to the consistency condition, tending to counteract the effects of small $T$.

We can assess the situation for the beam energy scan plasmas by focusing on the situation near to the quark matter critical point\footnote{The slope of the phase line \cite{kn:new,kn:newer,kn:newest} is thought to be negative but very small in magnitude; this is what we mean by saying that plasma temperatures significantly lower than that of the critical point are accompanied by very large values of the baryonic chemical potential. We therefore think it likely that if the consistency condition were to fail in the beam energy scan plasmas, it would do so near to the critical point, wherever that may be.} \cite{kn:mohanty,kn:satz,kn:race,kn:karsch}. A ``mainstream'' estimate of the location of this point might put it at ($T \approx 145$ MeV, $\mu_B \approx 300$ MeV) or ($T \approx 0.74$ fm$^{-1}$, $\mu_B \approx 1.52$ fm$^{-1}$), that is, at a temperature considerably lower than that of the RHIC plasmas, but at a much higher value of $\mu_B$. As we saw earlier when we briefly mentioned the LHC plasmas, the magnetic field drops more rapidly than the temperature at lower collision energies; let us be very conservative and assume that the maximal magnetic field in a peripheral collision producing a plasma near to the critical point is around $eB \approx 7.5 \times m_{\pi}^2$, or $B \approx 12.48$ fm$^{-2}$ (down from $eB \approx 10 \times m_{\pi}^2$ in the RHIC case), though we stress that this is almost certainly an over-estimate.

We find that, even though the temperature here is significantly lower than the 220 MeV assumed for the RHIC plasma temperature, the inequality (\ref{H}) is satisfied in this case; as predicted, a large baryonic chemical potential, assisted by a lower magnetic field, tends to ensure that the consistency condition is satisfied: see Figure 2.
\begin{figure}[!h]
\centering
\includegraphics[width=0.65\textwidth]{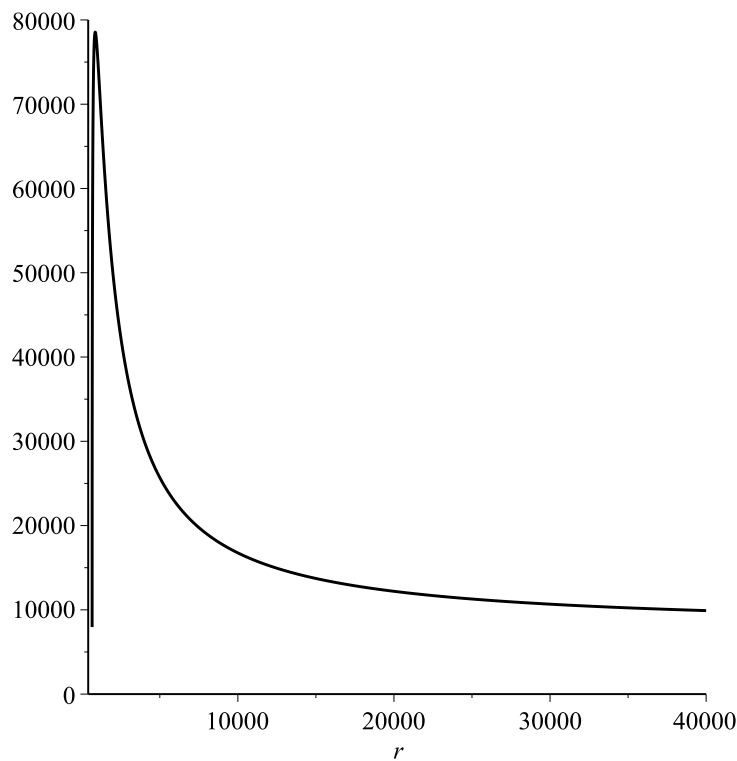}
\caption{$\mathfrak{S^{\m{E}}}(\m{AdSdyRN^{0}_{4}})(r)$, $T \approx 0.74$ fm$^{-1}$, $\mu_B \approx 1.52$ fm$^{-1}$, $B \approx 12.48$ fm$^{-2}$.}
\end{figure}

However, the location of the critical point is disputed; some recent works locate it far from this point in the quark matter phase diagram. To take two dramatically different estimates: the authors of \cite{kn:blaschke} have considered a value for the temperature of the critical point as low as $\approx$ 70 MeV $\approx 0.36$ fm$^{-1}$ (with an associated $\mu_B \approx$ 325 MeV $\approx 1.65$ fm$^{-1}$). Again we will be very conservative and estimate the maximal magnetic field in this case as $eB \approx 5 \times m_{\pi}^2$, or $B \approx 8.32$ fm$^{-2}$. Despite the extremely low temperature, we find that (\ref{H}) is again satisfied here: see Figure 3.
\begin{figure}[!h]
\centering
\includegraphics[width=0.75\textwidth]{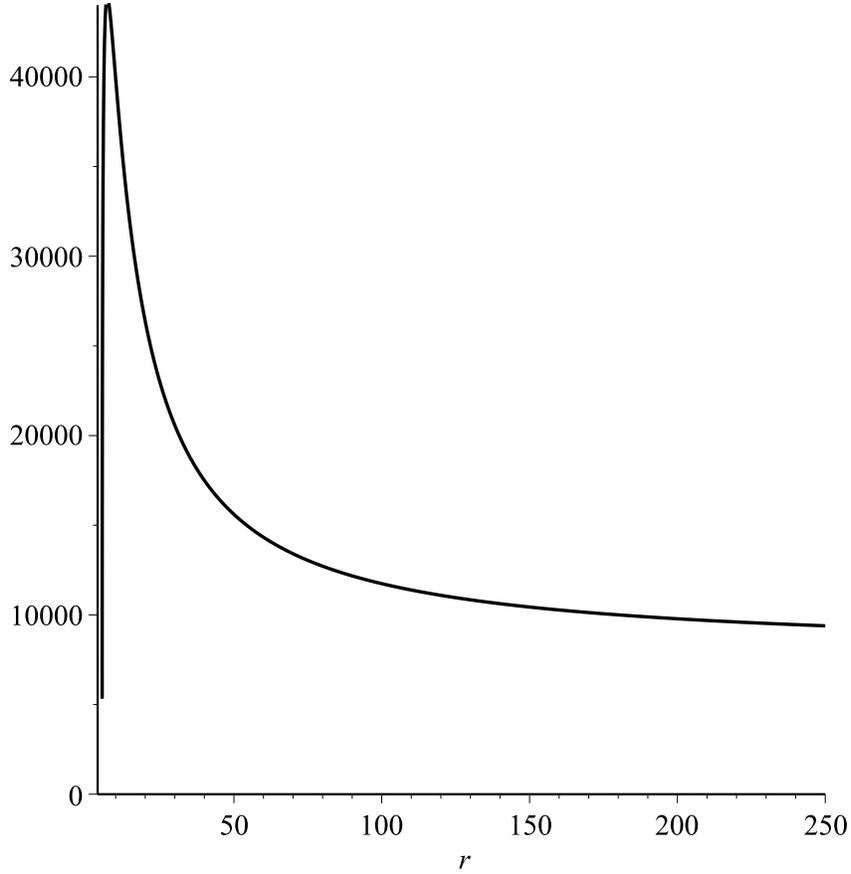}
\caption{$\mathfrak{S^{\m{E}}}(\m{AdSdyRN^{0}_{4}})(r)$, $T \approx 0.36$ fm$^{-1}$, $\mu_B \approx 1.65$ fm$^{-1}$, $B \approx 8.32$ fm$^{-2}$.}
\end{figure}

At the other extreme, the authors of \cite{kn:stony} put the critical point at around (T $\approx$ 165 MeV $\approx 0.84$ fm$^{-1}$, $\mu_B \approx$ 95 MeV $\approx 0.48$ fm$^{-1}$); but if we take once more $eB \approx 7.5 \times m_{\pi}^2$, or $B \approx 12.48$ fm$^{-2}$, we find yet again that the consistency condition is satisfied: see Figure 4.
\begin{figure}[!h]
\centering
\includegraphics[width=0.65\textwidth]{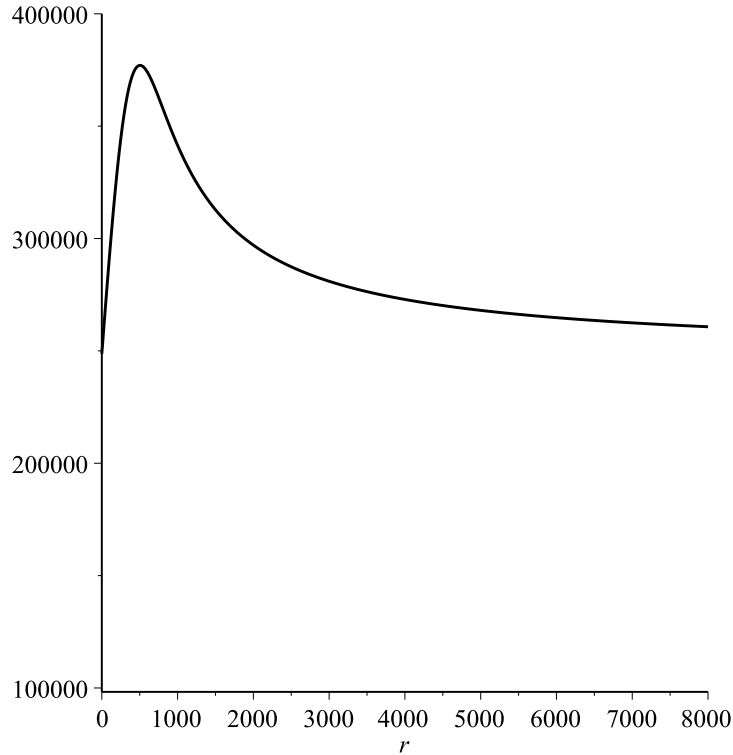}
\caption{$\mathfrak{S^{\m{E}}}(\m{AdSdyRN^{0}_{4}})(r)$, $T \approx 0.84$ fm$^{-1}$, $\mu_B \approx 0.48$ fm$^{-1}$, $B \approx 12.48$ fm$^{-2}$.}
\end{figure}

To summarize: with the new estimates of the largest possible magnetic fields produced in peripheral collisions, the RHIC plasma is in grave danger of violating the condition for a holographic dual to exist; the plasmas studied in the allied beam energy scan experiments are not.

This discussion neglects, of course, angular momentum, and our claim is that this is why there is an apparent conflict with the (actual or anticipated) data for certain low-$\mu_B$ peripheral collisions. We now consider an appropriately expanded version of the ``holographic dictionary'' and investigate the consequences.

\addtocounter{section}{1}
\section* {\large{\textsf{5. Angular Momentum Saves The Day I: Shear }}}
We will argue in this section that including (shearing) angular momentum allows us, in some important cases, to prevent violations of the fundamental inequality (\ref{A}). We begin with theoretical considerations, then turn to the specific case of the plasmas produced in the RHIC experiment.

\subsubsection*{{\textsf{5.1 Fixing the Parameters: Theory}}}

There are essentially two ways \cite{kn:KelvinHelm} in which angular momentum can be transferred to the QGP in a peripheral collision: as vorticity \cite{kn:viscous,kn:csernairecent1,kn:csernairecent2,kn:nagy,kn:nacs,kn:yin,kn:vortical}, or as shear \cite{kn:liang,kn:bec,kn:huang,kn:deng}. We begin with an exploration, using the gauge-gravity duality, of the consequences of including the latter. The former will be treated in the succeeding section.

Shear angular momentum can be studied holographically because there exist asymptotically AdS black holes (special cases of the Pleba\'nski--Demia\'nski family of metrics \cite{kn:plebdem,kn:grifpod}) which, through the frame-dragging associated with their angular momentum, induce a shearing effect at infinity. This has been studied in \cite{kn:77,kn:shear,kn:86}, to which we refer the reader for the details.

The basic quantity needed to specify the shearing in the plasma is the \emph{velocity profile} $v(x)$, which gives the velocity of the QGP, along the collision axis $z$, as a function of the transverse coordinate $x$. The shape of this function is determined by physics of the collision in a complex way described in \cite{kn:bec}\cite{kn:KelvinHelm}. It rises from zero along the symmetry axis to some maximum $V$ at the boundary of the collision zone; a typical shape (arising naturally from the specific bulk geometry we consider) is shown\footnote{The precise shape of the graph depends not just on $V$ but also on $L$. In order to obtain a reasonable shape, we choose $L \approx 10$ fm. See \cite{kn:shear}.} in Figure 5 (where for illustrative purposes we have taken $V \approx 1$).
\begin{figure}[!h]
\centering
\includegraphics[width=0.65\textwidth]{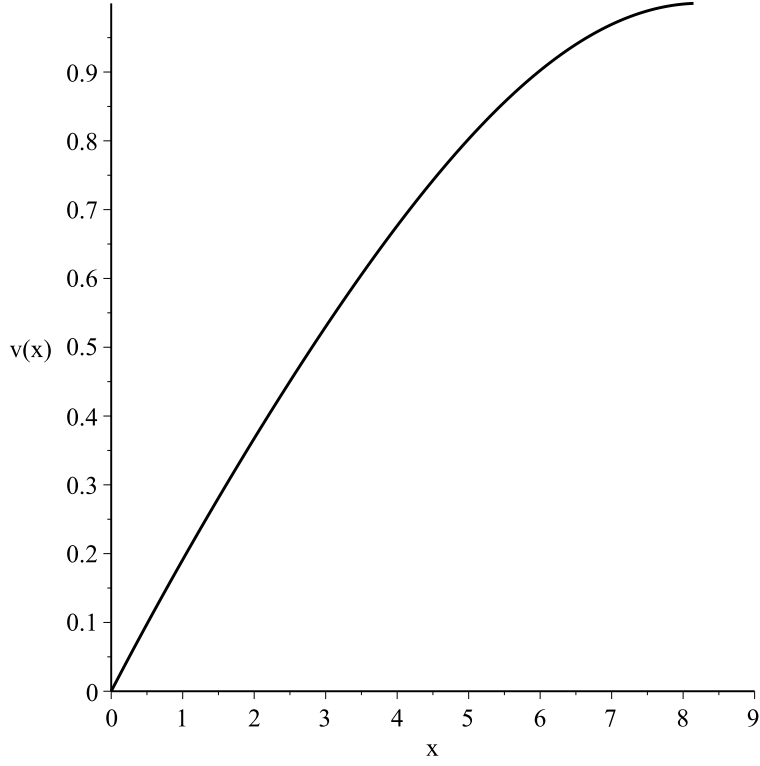}
\caption{A shearing velocity profile with $V=1$.}
\end{figure}

The metric we need is a generalization of the dyonic black hole metric given in equation (\ref{B}) above (again, with a topologically planar event horizon). In \cite{kn:shear} we found that, in order to obtain a velocity profile like the one in Figure 5, we needed to incorporate a parameter $\ell$ analogous to NUT charge; it proves to have a clear physical interpretation in this case, discussed below. These dyonic metrics generalize the ``KMV$_0$'' metrics given by Klemm, Moretti, and Vanzo \cite{kn:klemm} as the first examples of ``rotating'' planar black holes; hence we call them the ``$\ell$dyKMV$_0$'' metrics. They solve the AdS$_4$ Einstein-Maxwell equations (with asymptotic curvature $-1/L^2$), and have the form

\be
\label{I}
g(\ell {\rm dyKMV}_0)=-\frac{\Delta_r\Delta_\psi\rho^2}{\Sigma^2}\,\dif t^2+\frac{\rho^2\dif r^2}{\Delta_r}+\frac{\rho^2\dif\psi^2}{\Delta_\psi}+\frac{\Sigma^2}{\rho^2}\left[\omega\dif t-\dif\zeta\right]^2,
\ee
where
\ba
\label{J}
\rho^2&=&r^2+(\ell-a\psi)^2\cr
\Delta_r&=&\frac{(r^2+\ell^2)^2}{L^2}-8\pi M^*r+a^2+4\pi \left[Q^{*2}+P^{*2}\right]\cr
\Delta_{\psi}&=&1+\frac{\psi^2}{L^2}(2\ell-a\psi)^2\cr
\Sigma^2&=&(r^2+\ell^2)^2\Delta_\psi-\psi^2(2\ell-a\psi)^2\Delta_r\cr
\omega&=&\frac{\Delta_r\psi(2\ell-a\psi)-a(r^2+\ell^2)\Delta_\psi}{\Sigma^2}.
\ea
The electromagnetic potential one-form outside the black hole is \cite{kn:86}
\begin{equation}\label{K}
A(\ell {\rm dyKMV}_0)\;=\;A_t\m{d}t\;+\;A_{\zeta}\m{d}\zeta,
\end{equation}
where
\begin{equation}\label{KK}
A_t\;= -\,\frac {Q^*r + P^*(\ell-a\psi)} {\rho^2L}+\frac{Q^*r_h+P^*\sqrt{\ell^2+aL}} {L(r_h^2+\ell^2+aL)}
\end{equation}
\begin{equation}\label{KKK}
A_{\zeta}=\frac{-Q^*r(2\ell-a\psi)\psi+P^*\left(\psi-\frac{\ell}{a}\right)(r^2+\ell^2)} {\rho^2L}-\frac{Q^*r_hL-P^*\frac{\sqrt{\ell^2+aL}}{a}(r_h^2+\ell^2)}{L(r_h^2+\ell^2+aL)}.
\end{equation}
The coordinates here are straightforward generalizations of those appearing in equation (\ref{B}): again, $\psi$ and $\zeta$ define, at infinity, the reaction plane coordinates:
\begin{equation}\label{L}
\m{d}x = \frac{L\,\m{d}\psi }{\sqrt{1+\frac{\psi^2}{L^2}(2\ell-a\psi)^2}}, \;\;\; \m{d}z = L\,\m{d}\zeta .
\end{equation}
The reaction plane metric is then of course $\m{d}x^2 + \m{d}z^2$.

The parameter $a$ has its usual interpretation \cite{kn:86} as the specific angular momentum of the black hole (angular momentum per unit mass or energy)\footnote{For technical reasons it was necessary in \cite{kn:shear} to take $a$ to be negative. For convenience we have adjusted the relevant formulae so that, in the present work, $a$ should be taken to be positive.}. It retains this interpretation at infinity, as the ratio of the angular momentum and energy densities in the plasma. One can show \cite{kn:shear} that $\ell$ is related to $a$ and to the maximal velocity $V$ (discussed above) very simply:
\begin{equation}\label{M}
\ell^2=\;VaL.
\end{equation}
Note that $\ell$ has the same units as $a$, that is, in the units we use here, length.

As before, the black hole parameters $M^*, Q^*,$ and $P^*$ have physical interpretations in terms of the mass per unit horizon area, and so on; the precise relations are discussed in detail in \cite{kn:86}, but we do not need them here.

We are now in a position to state the generalized versions of equations (\ref{C}), (\ref{D}), (\ref{E}), and (\ref{F}) above: they take the form
\begin{equation}\label{N}
T\;=\;{K_V(r_h^2 + \ell^2)\over \pi r_hL^2}\;-\;{2K_VM^*\over r_h^2},
\end{equation}
\begin{equation}\label{O}
\mu_B/3\;=\;\frac{Q^*r_h+P^*\sqrt{\ell^2+aL}} {L(r_h^2+\ell^2+aL)},
\end{equation}
\begin{equation}\label{P}
B_m\;=\;{P^*\over L^3}\,J_V,
\end{equation}
\begin{equation}\label{Q}
\frac{(r_h^2+\ell^2)^2}{L^2}-8\pi M^*r_h+a^2+4\pi \left[Q^{*2}+P^{*2}\right]\;=\;0.
\end{equation}
Here $B_m$ denotes the spatial mean of the magnetic field; we use this to approximate the field in the model, which varies slowly with transverse position. We assume that $B_m$ corresponds at any given point to the actual magnetic field at that point\footnote{The actual field also varies slowly \cite{kn:ferrer} across the plasma, but we do not attempt to model this variation. If one wishes to consider the global variation of the magnetic field, then one should use $B_m$ to model the \emph{maximal} field, which is the field along the $x = 0$ axis.}. Finally, $K_V$ and $J_V$ are dimensionless constants, depending only on $V$, defined by
\begin{equation}\label{R}
K_V\;=\;\int_0^1{\m{d}p\over \sqrt{1+V^2p^2(2-p)^2}},
\end{equation}
\begin{equation}\label{S}
J_V = \int_0^1\sqrt{1+V^2p^2(2-p)^2}\;\m{d}p.
\end{equation}
The (somewhat intricate) derivations may again be found in \cite{kn:86}.

As before, we are now in a position, given the boundary data $T$, $\mu_B$, $B_m$, $a$, and $V$, together with $L$, to compute $\ell$ using equation (\ref{M}), and then we can solve these four equations for the four black hole parameters $M^*$, $Q^*$, $P^*$, and $r_h$.

By examining equation (\ref{K}), one sees that to obtain the Euclidean version of the metric given by equations (\ref{I}) and (\ref{J}), we need to complexify $a$, $\ell$, and $Q^*$, but not $P^*$. In particular, this means that we have a Euclidean version of $\Delta_r$, defined by
\begin{equation}\label{T}
\Delta_r^{\m{E}}\;=\;\frac{(r^2-\ell^2)^2}{L^2}-8\pi M^*r-a^2+4\pi \left[-Q^{*2}+P^{*2}\right].
\end{equation}
As before, we then have a ``Euclidean event horizon'', located at $r = r_h^{\m{E}}$, where $\Delta_r^{\m{E}}(r_h^{\m{E}}) = 0$. It is easy to show that $r_h^{\m{E}}$ always exists if $r_h$ does, that is, if the temperature is positive (so that cosmic censorship holds). Since we know how to compute $\ell$, $M^*$, $Q^*$, and $P^*$ from given boundary data, we can solve (\ref{T}) to determine $r_h^{\m{E}}$ from those data.

We can now compute $\mathfrak{S^{\m{E}}}(\ell {\rm dyKMV}_0)(r)$: it is given, for $a \neq 0$ (see \cite{kn:shear} for the relevant techniques), as usual up to a positive constant, by
 \begin{eqnarray}\label{U}
& &\mathfrak{S^{\m{E}}}(\ell {\rm dyKMV}_0)(r)\;=\;\sqrt{\Delta_r^{\m{E}}}\left(\sqrt{(\ell^2+aL)(r^2-\ell^2-aL)}+r^2\arcsin\frac{\sqrt{\ell^2+aL}}{r}\right) \nonumber \\
& & \;\;\;\;\;\;\;\;\;\;\;\;\;\;\;\;\;\;\;\;\;\;\;\;\;\;\;\;\;\;\;\;\;-\frac{2}{L}\sqrt{\ell^2+aL}\,\left[r^3-(r_h^{\m{E}})^3-(r-r_h^{\m{E}})(\ell^2+aL)\right].
\end{eqnarray}
As explained above, all of the constants in this expression are known or can be computed, given the boundary data $T$, $\mu_B$, $B_m$, $a$, $V$, and $L$. Thus, the precise form of this function is now known, in principle. (In practice it can only be determined numerically.)

In general terms, the graph of this function, for $a \neq 0$, takes the following shape. As always, it vanishes at the origin (the Euclidean event horizon) and then increases. It reaches a maximum, and then decreases to a minimum, at $r=r^{*}$; it then rises indefinitely; in fact asymptotically it approaches a straight line with positive slope. Therefore, the consistency condition (\ref{A}) is satisfied if and only if we have
\begin{equation}\label{UU}
\mathfrak{S^{\m{E}}}(\ell {\rm dyKMV}_0)(r^{*})\; \geq \; 0.
\end{equation}
This is the condition we need to check; it generalizes the inequality (\ref{H}) to the case with shearing angular momentum. Unfortunately, it does not seem to be possible to express (\ref{UU}) in an explicit form in the general case; in practice, the only way to check this condition is simply to examine the graph of $\mathfrak{S^{\m{E}}}(\ell {\rm dyKMV}_0)(r)$, to determine whether the minimum is indeed non-negative.

If $Q^* = P^* = 0$, then the metric here is an Einstein metric, and the Yamabe invariant at infinity is non-negative (see \cite{kn:shear}), so $\mathfrak{S^{\m{E}}}(\ell {\rm dyKMV}_0)(r)$ is never negative in that case, by the results of \cite{kn:lee,kn:wang,kn:wang2}; and so the consistency condition is satisfied. However, we know that this is not so when $Q^*$ is zero and $a$ is negligible, if $P^*$ is such that the inequality (\ref{AA}) is violated. The question is: what happens when $a$ is not negligible, as is the case for a real plasma associated with a large magnetic field?

The clearest answer to this question is found by examining the situation with reasonably realistic data, first from the RHIC experiment, then from the LHC.

\subsubsection*{{\textsf{5.2 Fixing the Parameters: RHIC Data}}}
We saw earlier that, with a typical RHIC temperature $T \approx 1.12$ fm$^{-1}$, zero baryonic chemical potential, and a somewhat high but quite possibly realistic estimate for the maximal magnetic field $eB \approx 10 \times m_{\pi}^2$, or $B \approx 16.64$ fm$^{-2}$, the graph of $\mathfrak{S^{\m{E}}}(\m{AdSdyRN^{0}_{4}})(r)$ (Figure 1) eventually crosses the horizontal axis. Let us now take the same values for $T$ and $B$, and use $\mathfrak{S^{\m{E}}}(\ell {\rm dyKMV}_0)(r)$ instead of $\mathfrak{S^{\m{E}}}(\m{AdSdyRN^{0}_{4}})(r)$.
To do so, we need an estimate\footnote{We retain $L = 10$ fm, and take $V \approx 1$ (since the velocity of the spectator nucleons is essentially that of light), for all numerical discussions henceforth. Note that $K_1 \approx 0.82473$ and $J_1 \approx 1.22991$.} for the maximal value of $a$, the ratio of the angular momentum density of the plasma to its energy density. Unfortunately this quantity is not known very precisely: see \cite{kn:86} for a discussion. There we settled on maximal values around $75$ fm, while acknowledging that substantially larger values may be possible. Fortunately, we will see that precise values for $a$ are not required for this purpose.

Numerical investigations reveal the following:

$\bullet$ The effect of including angular momentum is always to \emph{reduce} any conflict with the consistency condition (\ref{A}): $\mathfrak{S^{\m{E}}}(\ell {\rm dyKMV}_0)(r)$ is frequently everywhere positive in situations where $\mathfrak{S^{\m{E}}}(\m{AdSdyRN^{0}_{4}})(r)$ takes on negative values, for otherwise identical parameter values.

$\bullet$ The effect of including $a$ is frequently dramatic. For example, with the above parameter values, a value of $a$ as low as $a = 1$ fm suffices to keep $\mathfrak{S^{\m{E}}}(\ell {\rm dyKMV}_0)(r)$ positive everywhere: see Figure 6. Thus, our uncertainty as to the precise value of $a$ is not important: even with absurdly low values, the desired effect is achieved.
\begin{figure}[!h]
\centering
\includegraphics[width=0.65\textwidth]{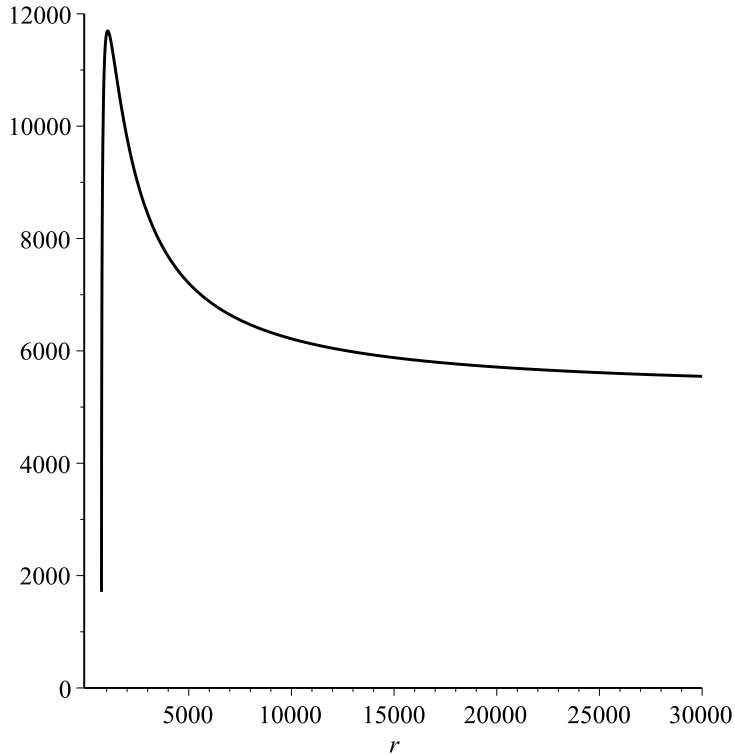}
\caption{$\mathfrak{S^{\m{E}}}(\ell {\rm dyKMV}_0)(r)$, $T \approx 1.12$ fm$^{-1}$, $\mu_B = 0$, $a = 1$ fm, $B \approx 16.64$ fm$^{-2}$.}
\end{figure}

In short, then, the situation regarding the RHIC data is clear-cut. If (shearing) angular momentum is neglected, then there is a severe risk that the consistency condition is violated by the maximal magnetic fields produced in peripheral collisions at that experiment; but when it is included, even though the precise corresponding values of $a$ are not known, all such risk is removed. \emph{In the RHIC regime, angular momentum dominates}.

However, the angular momentum need not take the form of shear: it could manifest itself as vorticity. Let us now consider that case.

\addtocounter{section}{1}
\section* {\large{\textsf{6. Angular Momentum Saves The Day II: Vorticity }}}
We now investigate the case of vorticity, in a way parallel to the argument of the preceding section. Although the technical details are surprisingly different, the overall strategy is the same, and the conclusions are more definite; so we shall be brief.

To study the holography of vorticity, we need an asymptotically AdS black hole such that the angular momentum induced at infinity by frame-dragging is independent of position there. The black hole we need turns out to be the AdS-Kerr-Newman black hole, with a topologically spherical (as opposed to planar) event horizon. (See \cite{kn:87} for a detailed discussion of this.)

In Boyer-Lindquist-like coordinates \cite{kn:cognola}, the dyonic AdS-Kerr-Newman metric takes the form
\begin{flalign}\label{V}
g(\m{AdSdyKN^{+1}_{4})} = &- {\Delta_r \over \rho^2}\Bigg[\,\m{d}t \; - \; {a \over \Xi}\m{sin}^2\theta \,\m{d}\phi\Bigg]^2\;+\;{\rho^2 \over \Delta_r}\m{d}r^2\;+\;{\rho^2 \over \Delta_{\theta}}\m{d}\theta^2 \\ \notag \,\,\,\,&+\;{\m{sin}^2\theta \,\Delta_{\theta} \over \rho^2}\Bigg[a\,\m{d}t \; - \;{r^2\,+\,a^2 \over \Xi}\,\m{d}\phi\Bigg]^2,
\end{flalign}
where the ``$+1$'' indicates the spherical topology of the event horizon, and where
\begin{eqnarray}\label{eq:W}
\rho^2& = & r^2\;+\;a^2\m{cos}^2\theta, \nonumber\\
\Delta_r & = & (r^2+a^2)\Big(1 + {r^2\over L^2}\Big) - 2Mr + {Q^2 + P^2\over 4\pi},\nonumber\\
\Delta_{\theta}& = & 1 - {a^2\over L^2} \, \m{cos}^2\theta, \nonumber\\
\Xi & = & 1 - {a^2\over L^2}.
\end{eqnarray}
The associated electromagnetic field potential (outside the event horizon) is given by
\begin{equation}\label{WW}
A = -\,{Q\,\Xi\, r\over 4\pi\rho^2}\left[\m{d}t-{a\,\m{sin}^2\theta\over \Xi}\m{d}\phi\right]-{P\,\Xi\,\m{cos} \theta\over 4\pi\rho^2} \left[a\,\m{d}t - {r^2+a^2 \over \Xi}\m{d}\phi\right].
\end{equation}

Here $L$ is the asymptotic curvature length scale, $a$ is the specific angular momentum as before, and the black hole parameters are related to physical parameters on the boundary in non-trivial ways, leading to the holographic dictionary for this case:
\begin{equation}\label{X}
T\;=\;{r_h \Big(1\,+\,a^2/L^2\,+\,3r_h^2/L^2\,-\,{a^2\,+\,\{Q^2+P^2\}/4\pi \over r_h^2}\Big)\over 4\pi (a^2\,+\,r_h^2)},
\end{equation}
\begin{equation}\label{Y}
\mu_B \,=\,{3\Xi\left(Q\,r_h+aP\right)\over 4\pi L\left(r_h^2+a^2\right)},
\end{equation}
\begin{equation}\label{Z}
B\,=\,{\Xi \,P\over L^3}.
\end{equation}
\begin{equation}\label{ALPHA}
\Delta_r(r_h)\;=\;(r_h^2+a^2)\Big(1 + {r_h^2\over L^2}\Big) - 2Mr_h + {Q^2 + P^2\over 4\pi}\;=\;0,
\end{equation}
the notation being as before.

The principal technical problem with using this metric is that the spatial sections at infinity have the topology of a two-sphere; one has to approximate the space near to one of the poles by the tangent plane there, and use that tangent plane as a model of the reaction plane (the customary $x - z$ plane) used to study heavy-ion collisions. To ensure this, one has to take $L$ to be larger than any of the other length scales (particularly, in view of the definition of $\Xi$, the length scale defined by $a$) in the problem; we will use $L = $ 100 fm when discussing the RHIC experiments (maximal value of $a \approx 75$ fm); a larger value would be required for a discussion of the LHC case, where $a$ can be substantially larger.

Up to an overall positive factor, $\mathfrak{S^{\m{E}}}(r)$ for this geometry takes the form \cite{kn:74}
\begin{eqnarray}\label{BETA}
\mathfrak{S^{\m{E}}}(\m{AdSdyKN^{+1}_{4})}(r) & = & \left\{r\sqrt{(r^2-a^2)\Big(1 + {r^2\over L^2}\Big) - 2Mr + {- Q^2 +P^2\over 4\pi}}\right.\,\,\times  \nonumber\\ &&\;\;\;\left.\Bigg[\sqrt{1-{a^2\over r^2}}+ {r\over a}\, \m{arcsin}{a\over r}\Bigg]\right\}-\;{2r^3\over L}\Bigg[1 - {a^2\over r^2}\Bigg] \nonumber \\ && \;\;\;+\;{2(r_h^E)^3\over L}\Bigg[1 - {a^2\over (r_h^E)^2}\Bigg],
\end{eqnarray}
where $r = r_h^E$ is defined as before.

Recall that we found, in the case of shearing angular momentum, that even an unrealistically small value of $a$ sufficed to prevent $\mathfrak{S^{\m{E}}}(\ell {\rm dyKMV}_0)(r)$ from becoming negative at any point (Figure 6). The same conclusion holds here: see Figure 7.

\begin{figure}[!h]
\centering
\includegraphics[width=0.65\textwidth]{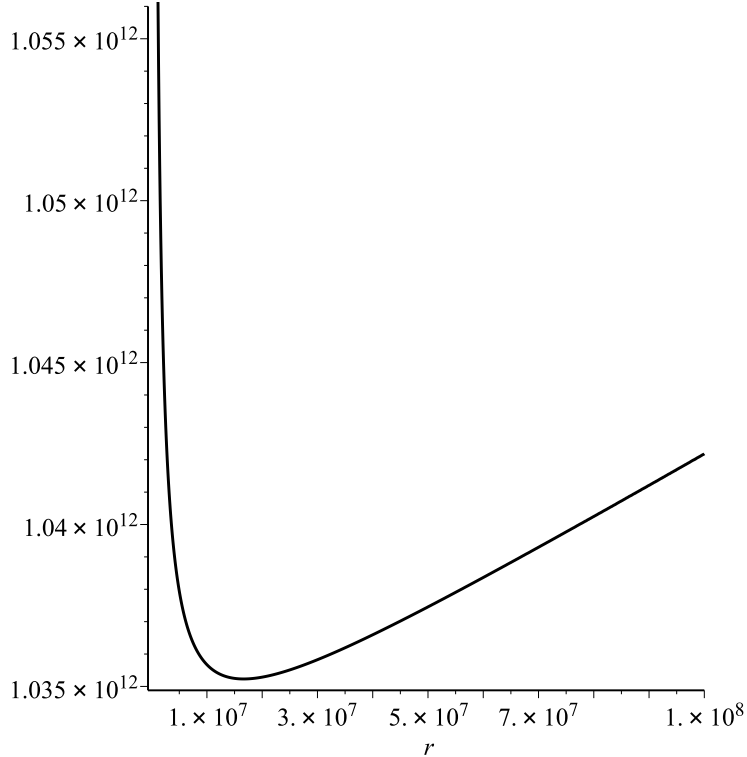}
\caption{$\mathfrak{S^{\m{E}}}(\m{AdSdyKN^{+1}_{4})}(r)$,$T \approx 1.12$ fm$^{-1}$, $\mu_B = 0$, $a = 1$ fm, $B \approx 16.64$ fm$^{-2}$.}
\end{figure}

In point of fact, we have been unable to find any combination of parameter values that causes $\mathfrak{S^{\m{E}}}(\m{AdSdyKN^{+1}_{4})}(r)$ to become negative at any value of $r$; this case is apparently never in conflict with the consistency condition. Thus, vorticity is very effective in restoring the consistency condition.

The angular momentum transferred to the actual plasma produced in peripheral heavy-ion collisions can take either form, shearing or vorticity, depending on the nature of the plasma itself. Our conclusion is that, in the case of the RHIC experiment (and the beam energy scans), both forms resolve the apparent conflict with the string-theoretic consistency condition: these plasmas do have a holographic dual, provided that one takes angular momentum into account.

The situation at the LHC experiments is very different, as we now show.

\addtocounter{section}{1}
\section* {\large{\textsf{7. The Case of the LHC Plasmas}}}
Heavy-ion collisions are of course also studied at the LHC, in the ALICE and other experiments \cite{kn:armesto}. The conditions in these experiments are considerably more extreme than in the RHIC, and, in view of the asymptotic freedom of QCD, it is by no means clear that the plasmas produced in them are sufficiently strongly coupled as to render a holographic treatment appropriate: perhaps perturbative methods are more suitable in this regime. (See for example \cite{kn:noodles} for the literature discussing this question; and see \cite{kn:quench} for recent evidence suggesting that the LHC plasmas may indeed differ in important ways from their RHIC counterparts.)

In particular, it has been found \cite{kn:barb} that the holographic jet quenching model that works rather well for RHIC data does not work well for LHC data. While it may be possible to remedy this \cite{kn:ficnar1,kn:ficnar2}, the alternative interpretation is that holography is simply not applicable in the LHC case. If that is so, then the question becomes: at what point, between the RHIC and LHC domains, does holography cease to be relevant?

Perhaps we should step back at this point and ask another question: even leaving aside doubts as to how strongly the LHC plasmas are coupled, could it be that holography cannot be applied to a generic plasma under such extreme conditions? Could it be, for example, that the field theories used to model the generic LHC plasmas do not have consistent bulk duals (because they are typically associated with even more enormous magnetic fields than their RHIC counterparts)? With this background, let us now investigate the status of string-theoretic consistency for the LHC plasmas.

In this case, $T$ is considerably larger than at the RHIC, $T \approx 300$ MeV $\approx 1.52$ fm$^{-1}$, and the baryonic chemical potential is even closer to zero. The maximal magnetic field in peripheral LHC collisions is however expected to be far larger than in the RHIC case: a typical recent estimate is cited in \cite{kn:denghuang,kn:gergend}, $eB \approx 70 \times m_{\pi}^2$ (over 1.3 GeV$^2$), or $B \approx 117$ fm$^{-2}$, definitely exceeding the right side of the inequality (\ref{AA}) in this case ($\approx 25.8$ fm$^{-2}$). (For a discussion of the difficulties of actually observing the effects of even such enormous fields, see \cite{kn:data}.)

However, the shearing angular momentum transferred to the plasma in this case is also very much larger than in the RHIC case \cite{kn:bec}. We saw in the case of the RHIC plasma that even an extremely low value for $a$, around 1 femtometre, sufficed to restore consistency. Here, the angular momentum density is around 14 times larger, but the energy density \cite{kn:aliceenergy} is around 2.3 times larger ---$\,$ recall that $a$ is the ratio of the angular momentum to the energy densities ---$\,$ so, if we accept $a \approx 75$ fm for the RHIC, then the value of $a$ associated with the maximal magnetic fields at the LHC is $a \approx 457$ fm. One would think that if $a = 1$ fm suffices to save consistency for the RHIC plasma, then surely a value over 450 times larger should perform the same service for the LHC plasma.

\emph{Surprisingly, however, that is not the case}. In the LHC regime, the effects of $B$ (tending to violate consistency) completely outstrip those of $a$ (tending to restore it), to the extent that even such a huge increase in $a$ fails to save the day: see Figure 8.
\begin{figure}[!h]
\centering
\includegraphics[width=0.65\textwidth]{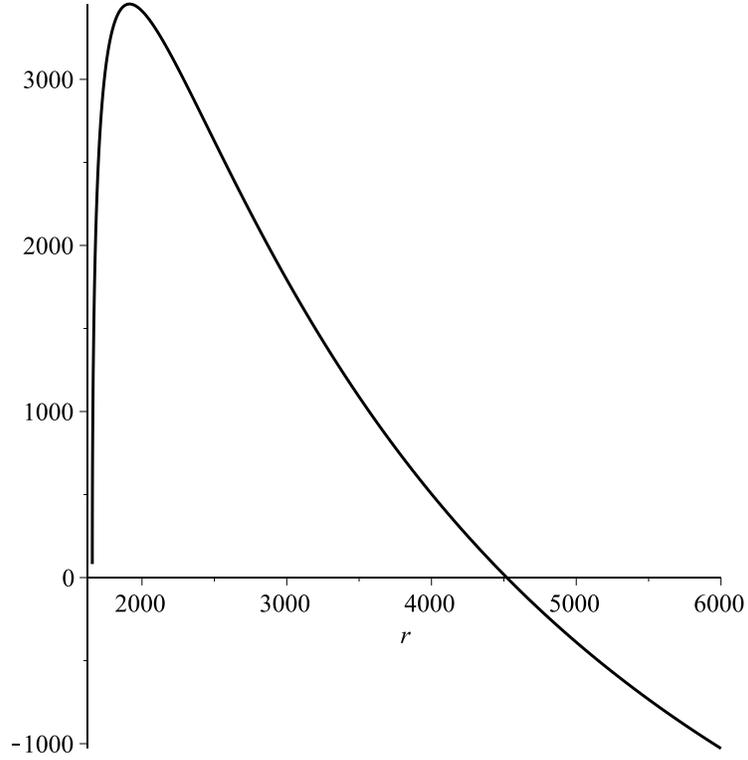}
\caption{$\mathfrak{S^{\m{E}}}(\ell \m{dyKMV}_0)(r)$, $T \approx 1.52$ fm$^{-1}$, $\mu_B = 0$, $a = 457$ fm, $B \approx 117$ fm$^{-2}$.}
\end{figure}
In fact, we are not even close to satisfying the consistency condition here: one has to take $a$ over 3000 fm, an utterly unrealistic value\footnote{Actually, 3100 fm suffices.}, to force $\mathfrak{S^{\m{E}}}(\ell \m{dyKMV}_0)(r)$ to be strictly positive, with these values of the temperature and magnetic field: see Figure 9.
\begin{figure}[!h]
\centering
\includegraphics[width=0.65\textwidth]{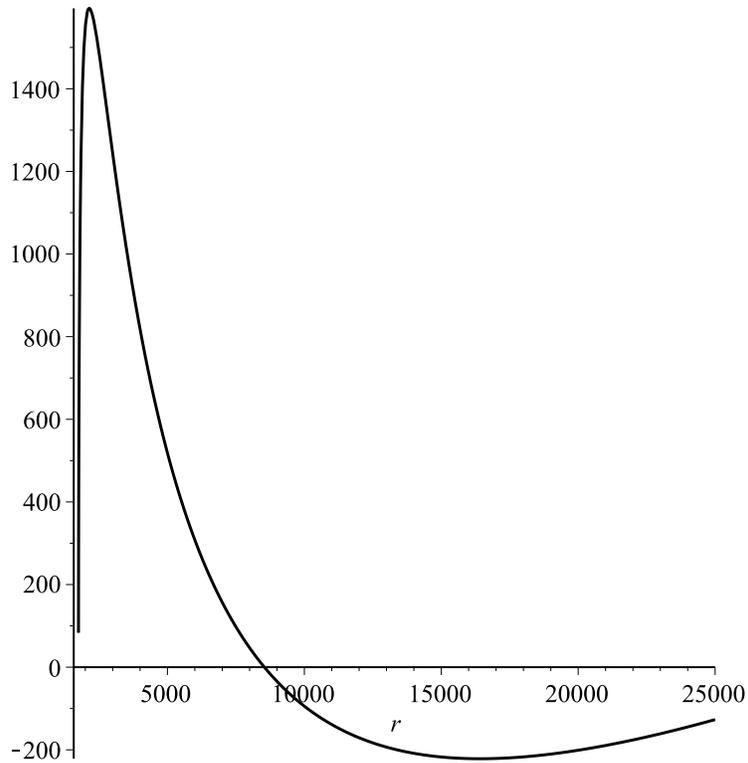}
\caption{$\mathfrak{S^{\m{E}}}(\ell \m{dyKMV}_0)(r)$, $T \approx 1.52$ fm$^{-1}$, $\mu_B = 0$, $a = 3000$ fm, $B \approx 117$ fm$^{-2}$.}
\end{figure}
This is important, since it means that the consistency condition is violated by the LHC plasma not only at the most extreme values of $B$, but also for much smaller values. In other words, a \emph{generic} peripheral collision at the LHC produces such a plasma, not just those with impact and other parameters finely tuned. To put it another way, only the most central, and the most extremely peripheral\footnote{See \cite{kn:bec} for a detailed discussion of the angular momentum transfer as a function of the impact parameter. A similar analysis holds for the magnetic field.}, LHC collisions avoid producing a plasma that violates the consistency condition, the inequality (\ref{A}). \emph{In the LHC regime, the magnetic field dominates.}

In general, the expectation (\cite{kn:linear} and references therein) is that the maximal magnetic fields in peripheral collisions grow roughly linearly
with the centre-of-momentum energy of the colliding nucleons. But the (squared) temperature grows much more slowly with the energy; thus, if consistency is generically violated at the LHC, one can expect\footnote{On the other hand, the angular momentum transfer \emph{also} grows approximately linearly with collision energy; however, as we have seen, at extremely high energies, angular momentum is completely ineffective in restoring the consistency condition.}, barring a major surprise, that it will be comprehensively violated at future facilities \cite{kn:SppC,kn:FCC1,kn:FCC2} involving heavy-ion collisions with energies in the range 20 - 40 TeV, producing plasmas with temperatures unlikely to be higher than about 400 MeV at the relevant point in the evolution of the plasma. (That is, the relevant magnetic fields could be around 7 times larger than those at the LHC, but the corresponding temperatures may be only around 30$\%$ higher.)

All this, of course, reinforces the contention that holographic methods should not be applied to the systems studied at the LHC or at future facilities of the same kind. The question, then, as mentioned above, is: how far above the RHIC temperatures and fields does one have to go in order to reach the regime in which holography is unlikely to be very useful?

This is a complicated question, but to obtain a rough estimate, let us interpolate linearly between the RHIC and LHC data (angular momentum density, energy density, temperature, magnetic field). If we do so, then we find that the consistency condition is satisfied if the RHIC collision energy is scaled up from 200 GeV to 350 GeV (T= 1.14 fm$^{-1}$, a = 122 fm, B = 29 fm$^{-2}$); see Figure 10.

\begin{figure}[!h]
\centering
\includegraphics[width=0.65\textwidth]{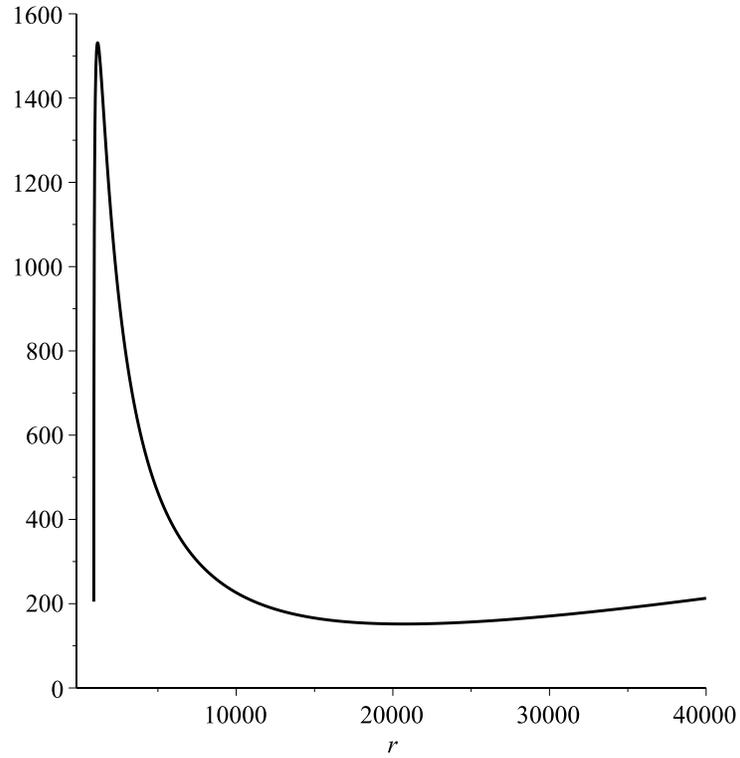}
\caption{$\mathfrak{S^{\m{E}}}(\ell \m{dyKMV}_0)(r)$, Collision energy $\sqrt{s_{NN}} \approx $ 350 GeV .}
\end{figure}

But if we go further, up to 400 GeV collision energy (T= 1.15 fm$^{-1}$, a = 136 fm, B = 33.3 fm$^{-2}$), still of course far below the LHC collision energy for heavy ions ($\sqrt{s_{NN}}$ = 2.76 TeV), we find that the consistency condition ceases to be satisfied: see Figure 11.

\begin{figure}[!h]
\centering
\includegraphics[width=0.65\textwidth]{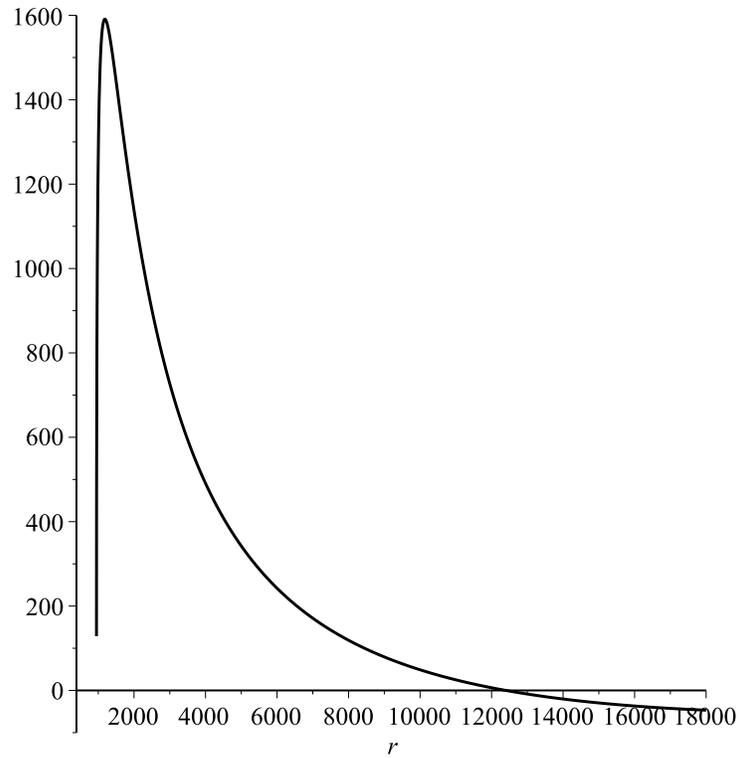}
\caption{$\mathfrak{S^{\m{E}}}(\ell \m{dyKMV}_0)(r)$, Collision energy $\sqrt{s_{NN}} \approx $ 400 GeV .}
\end{figure}

The boundary\footnote{Notice once again that the effect of including the angular momentum is to maintain holographic consistency up to considerably higher values of $B/T^2$ than would otherwise be possible. For example, inequality (\ref{AA}) forbids any values of $B/T^2$ beyond about 11.14; but the model, including angular momentum, with $\sqrt{s_{NN}} \approx $ 350 GeV, allows $B/T^2 \approx $ 22.3. This is of interest in connection with recent works \cite{kn:bigB,kn:romulorouge,kn:gg} which consider situations in which $eB \gg T^2$. } evidently lies between the collision energies of the RHIC and LHC experiments, but much closer to the former than to the latter; roughly, somewhere between 350 and 400 GeV.

Of course, it may be that, just as the inclusion of angular momentum solves the problem of the apparent non-existence of a holographic dual in the case of the RHIC plasma, so also the inclusion of some other neglected effect will ``save the day'' in the present case. That effect will need to be strong indeed to raise the threshold from (say) just over 350 GeV to 2.76 TeV. (However, there are some subtleties here, discussed briefly in the Conclusion.)

\addtocounter{section}{1}
\section* {\large{\textsf{8. Conclusion}}}
That string theory is a very rigid structure is a well-rehearsed idea, yet fully explicit instances are few. The work of Ferrari and Rovai \cite{kn:ferrari1,kn:ferrari2,kn:ferrari3,kn:ferrari4} has the virtue of leading directly to very explicit constraints which must be satisfied if a field theory is to have a holographic dual. In this work, we have attempted to apply (just) one of their constraints to the plasmas produced in collisions of heavy ions. The results are easily summarized:

$\bullet$ Among all of the effects arising in such collisions, the \emph{extreme magnetic fields} experienced by plasmas produced in peripheral collisions pose the strongest challenge to the existence of a bulk dual.

$\bullet$ The high values of the baryonic chemical potential characteristic of the plasmas in the beam energy scan experiments \cite{kn:STAR,kn:shine,kn:nica,kn:fair,kn:BEAM,kn:luo} are nevertheless able to counteract the effect of the magnetic fields: there is no difficulty in applying holography to this case.

$\bullet$ In the low-$\mu_B$ plasmas produced at the RHIC, there is a serious danger that the consistency condition will be violated. This danger is eliminated, however, when one recalls that the magnetic fields are invariably accompanied by very high angular momentum densities.

$\bullet$ In the case of the LHC plasmas, however, if the angular momentum arises from an shearing motion in the plasma, then it does \emph{not} suffice to overcome the effect of the magnetic field.

$\bullet$ Using a simple linear scaling, one finds that the transition between the RHIC and LHC results occurs much closer to the RHIC case, probably at a collision energy below 400 GeV. It will therefore be difficult, though perhaps not impossible, to establish the existence of a holographic dual of the LHC plasmas, by including still another effect neglected here; though we do not know what that effect might be.

In summary: the stringy account of the bulk physics is most clearly internally consistent in the case of the RHIC plasmas and those to be explored more fully in the beam energy scans. Perhaps it is towards large $\mu_B$ (and angular momentum density) that gauge-gravity investigations are best directed.

We close with two comments relating to future work. The first point is that we stress that the consistency condition we have used here is just one aspect of the complex investigated in \cite{kn:ferrari1,kn:ferrari2,kn:ferrari3,kn:ferrari4}. It is quite possible that other, equally restrictive conditions are imposed on boundary field theories by the simple demand that their ostensible bulk duals actually exist in string theory. Clearly, expressing these restrictions explicitly, and interpreting them in terms of heavy ion data, is a project of great importance.

The second point relates to the way angular momentum restored the consistency condition in the case of the RHIC plasmas. It is very remarkable that the amount of specific angular momentum required to do so is so small; the necessary value of $a$ is at least two orders of magnitude smaller than realistic values for these collisions. To put it another way: in a sense, the situation pictured in Figure 1 arises because $a$ has been \emph{fine-tuned to zero} --- if we had begun with a ``more generic'' bulk geometry (with the metric in (\ref{I}) instead of the one in (\ref{B})), then almost any value of $a$ would have avoided the problem. This prompts the question: could it be that the problem regarding the LHC plasmas arises in precisely the same way, because we are neglecting some parameter (effectively fine-tuning it to zero)? As we know, the consistency condition is not close to being satisfied by the LHC plasmas, but, as the dependence on the parameters is evidently delicate, we should be cautious here.

Of course it is true that, as in any holographic model, we are indeed neglecting a host of lower-order effects: to name but two, we are ignoring the possible coupling of the magnetic field with other fields that may be present in the bulk (most notably, with dilatonic fields, but also with the probe branes used to derive (\ref{A}) in the first instance), and we are neglecting possible effects due to the expansion of the plasma. In view of this, the question becomes: are our results in Section 7 robust against perturbations of the bulk geometry, corresponding to the effects we have neglected?

This is a difficult question, but we can begin to approach it by means of a controlled deformation of the bulk induced by considering a dilatonic field coupled to the magnetic field. The great advantage of this is that one can introduce such a field into the bulk without disturbing the asymptotically AdS character of the bulk geometry (by carefully choosing the dilaton potential). One can then study the effect on the quantity $\mathfrak{S^{\m{E}}}$ as the dilaton-magnetic coupling $\alpha$ is varied. We regard this as a sort of probe of the bulk geometry, to detect whether (for given values of the physical parameters, for example, for $T$ and $B$ fixed at typical LHC values) the condition (\ref{A}) is immediately restored by very small values of $\alpha$, just as it was restored in the RHIC case by very small values of $a$.

The computations are somewhat intricate, and will be reported elsewhere; in summary, however, we have found that this does \emph{not} happen. We find that (\ref{A}) can indeed be restored by such a deformation, but that the required value of the coupling $\alpha$ is (in a sense that can be defined rather precisely) \emph{large}. Thus we have some evidence that the results of Section 7 are indeed robust. We will return to this in the near future.

\addtocounter{section}{1}
\section*{\large{\textsf{Acknowledgements}}}
The author wishes to acknowledge helpful discussions with Prof Soon Wanmei, and with Jude and Cate McInnes.

\end{document}